\documentclass[conference]{IEEEtran}
\IEEEoverridecommandlockouts

\usepackage{setspace}
\usepackage[T1]{fontenc}
\usepackage{color,array,amsthm}
\usepackage{graphicx}
\usepackage{cite}
\usepackage{amsmath,amssymb,amsfonts}
\usepackage{textcomp}
\usepackage{xcolor}
\usepackage{colortbl}
\usepackage{url}
\usepackage{multirow}
\usepackage{graphicx}
\usepackage{wrapfig}
\usepackage{latexsym}
\usepackage{algorithm}
\usepackage{algpseudocode}
\usepackage{subcaption}
\usepackage[top=0.70in, bottom=1.05in, left=0.64in, right=0.64in]{geometry}
\def\BibTeX{{\rm B\kern-.05em{\sc i\kern-.025em b}\kern-.08em
    T\kern-.1667em\lower.7ex\hbox{E}\kern-.125emX}}
\begin{document}


\title{Green Satellite Networks Using Segment Routing and Software-Defined Networking \\

}
\author{
\IEEEauthorblockN{
Jintao Liang\IEEEauthorrefmark{1},
Pablo G. Madoery\IEEEauthorrefmark{1},
Chung-Horng Lung\IEEEauthorrefmark{2},
Halim Yanikomeroglu\IEEEauthorrefmark{1},
Gunes Karabulut Kurt\IEEEauthorrefmark{1}\IEEEauthorrefmark{3}
}
\vspace*{0.3cm}
\IEEEauthorblockA{\IEEEauthorrefmark{1} Non-Terrestrial Networks (NTN) Lab, Department of Systems and Computer Engineering, Carleton University, Canada}
\IEEEauthorblockA{\IEEEauthorrefmark{2}Department of Systems and Computer Engineering, Carleton University, Canada}
\IEEEauthorblockA{\IEEEauthorrefmark{3} Poly-Grames Research Center, Department of Electrical Engineering, Polytechnique Montréal, Montréal, Canada}
}
\maketitle

\begin{abstract}
This paper presents a comprehensive evaluation of network performance in software defined networking (SDN)-based low Earth orbit (LEO) satellite networks, focusing on the Telesat Lightspeed constellation. We propose a green traffic engineering (TE) approach leveraging segment routing IPv6 (SRv6) to enhance energy efficiency. Through simulations, we analyze the impact of SRv6, multi-protocol label switching (MPLS), IPv4, and IPv6 with open shortest path first (OSPF) on key network performance metrics, including peak and average CPU usage, memory consumption, packet delivery rate (PDR), and packet overhead under varying traffic loads. Results show that the proposed green TE approach using SRv6 achieves notable energy efficiency, maintaining lower CPU usage and high PDR compared to traditional protocols. While SRv6 and MPLS introduce slightly higher memory usage and overhead due to their advanced configurations, these trade-offs remain manageable.  Our findings highlight SRv6 with green TE as a promising solution for optimizing energy efficiency in LEO satellite networks, contributing to the development of more sustainable and efficient satellite communications.
\end{abstract}


\section{Introduction}

The rapid development of low Earth orbit (LEO) satellite networks represents a crucial advance in achieving global connectivity in recent years. LEO networks are beneficial for providing high-speed internet access to remote regions, such as the northern part of Canada. Several leading corporations, including SpaceX's Starlink\cite{starlink_fcc} and Telesat's Lightspeed\cite{lightspeed_fcc}, are deploying their own LEO satellite constellations. 

Simultaneously, software-defined networking (SDN) has transformed network management by leveraging software-based controllers and separating the control plane from the data plane\cite{sdn_basis}. This separation adds flexibility, enabling scalable and responsive networks in dynamic traffic environment. 

Due to the growing number of devices and traffic volume, green traffic engineering (TE) approaches have emerged as crucial strategies to reduce energy consumption \cite{green_networking_survey}. Implementing green TE in LEO satellite networks is increasingly necessary to create energy-efficient and long-lived networks. Green TE aims to minimize energy consumption by optimizing resource utilization, reducing CPU usage, and implementing intelligent power management strategies. 

On the other hand, segment routing IPv6 (SRv6) simplifies the control plane by using source routing, eliminating the need for complex routing while maintaining efficient data forwarding\cite{sr architecture}. With SDN, the network topology can be continuously monitored, allowing dynamic path adjustments and resource optimization based on real-time conditions. SRv6 can facilitate fast rerouting and efficient TE by enabling the source node to specify the route through the network \cite{sr reroute} \cite{sr te}. The integration of SDN with SRv6 has shown considerable promise in enabling energy-efficient approaches \cite{jacob sr}. This combination is particularly suitable for energy-efficient management in LEO satellite networks, where optimizing energy use is critical to prolong the lifetime of the satellites\cite{FSO_Satellite_Networks}. 

This paper investigates the integration of green TE and SRv6 within LEO satellite networks. We simulate the Telesat Lightspeed constellation and evaluate key performance metrics. In evaluating different routing protocols, such as IPv4 and IPv6 based on open shortest path first (OSPF), multi-protocol label switching (MPLS), and SRv6, across performance metrics like peak and average CPU usage, memory usage, packet delivery rate (PDR) and packet overhead, distinct patterns emerge. Original SRv6 consistently demonstrates a balanced approach with a lower peak and average CPU usage compared to others, while maintaining a higher PDR as traffic load increases. The green TE proposed using SRv6 shows even better performance in terms of average CPU usage. These insights reveal that the proposed green TE, despite a slightly higher packet overhead and memory usage, provides a favorable trade-off by reducing computational load and thereby CPU usage due to green rerouting based on energy usage. This balance is crucial for LEO satellites, where energy and computational resources are limited, and suggests that SRv6, particularly in green TE, may be optimal for SDN-based satellite networks aiming for sustainable and high-performance communication. The contributions of this work are as follows:
\begin{itemize}
    \item A green TE approach leveraging SRv6 to reduce CPU usage and lower energy consumption in LEO satellites without compromising network performance.
    \item A performance evaluation of routing protocols under varying traffic loads in the Lightspeed constellation.
    \item Practical insights into managing green TE in LEO satellite networks, along with a discussion of key research challenges for future work.
\end{itemize}


The remainder of the paper is organized as follows: Section II covers related work and motivation. Section III presents the system model and methodology employed. Section IV details the simulation setup, analyzes performance results, and provides key insights. Finally, Section V summarizes the conclusions and outlines directions for future research.

\section{Motivation and Related Work}
Several studies discuss the integration of SDN in satellite networks, SDN with SRv6, and green traffic engineering (TE), as seen in \cite{sdn_placement,sdn_icn,sdn_leo, energy eff sat, green sat, dynamic_links, srv6_overview, green_traffic_engineering, sr and green sdn, TE_IP_Routing}. In \cite{sdn_placement}, the authors study the effects of traffic variability on the average flow setup time. They introduce an architecture for a SDN-based LEO satellite constellation and examine data and control planes partitioning within a dynamically changing satellite topology. 

In \cite{sdn_icn}, the authors propose a satellite network architecture to enhance network management by decoupling the forwarding and control functions in SDN. The paper addresses the challenges related to the inefficiency and complexity of conventional satellite network routing algorithms. In \cite{sdn_leo}, the authors explore the integration of SDN with LEO satellite networks. They propose a centralized control system featuring a master controller and several sub-controllers, which facilitates a routing algorithm designed for efficient traffic management.

In \cite{energy eff sat}, the authors examine energy efficiency in satellite networks, highlighting opportunities for optimization across various layers and components, from transmission power adjustments to resource allocation strategies. In \cite{green sat}, the authors propose several routing algorithms, including green routing, which can balance energy efficiency with path length and link utilization. The study shows that the proposed algorithms can significantly extend the lifetime of the satellite battery by managing power more effectively and selectively placing satellites in sleep mode. In \cite{dynamic_links}, the authors address the energy inefficiency of keeping idle laser inter-satellite links (LISLs) active in LEO constellations by proposing on-demand routing with dynamic LISL setup. They model the impact of setup delay as a latency penalty, formulate a nonlinear optimization problem, and propose three heuristic algorithms balancing performance and complexity. 

In \cite{srv6_overview}, the authors present SRv6 as a promising source‐routing protocol for communications. They explain how SRv6 integrates with existing IPv6 infrastructure and discuss standardization efforts. In \cite{green_traffic_engineering}, the authors investigate green TE in SDN using SR, proposing idle mode to reduce energy consumption. Results demonstrate SR’s potential to balance energy efficiency with network performance. In \cite{sr and green sdn}, the authors examine how SR features simplifies traffic management and conclude that the proposed integration offers a holistic approach to green TE that improves both energy efficiency and network resilience. In \cite{TE_IP_Routing}, the authors present a TE approach for IP networks using traditional protocols such as OSPF, proposing a framework that measures topology and traffic demands, and controls routing via link weight adjustments. 

Building on prior research, numerous works have studied SDN and green TE in various network contexts. However, this work stands out by proposing an innovative satellite network architecture that integrates green TE with SRv6. Unlike previous studies, our research introduces a comparative analysis of green TE using SRv6 against traditional routing protocols—MPLS, IPv6, and IPv4—within the Lightspeed constellation. Through performance evaluation, we offer new insights into the operational efficiency of these protocols in LEO satellite networks.

To the best of our knowledge, this is the first study to investigate key performance metrics of various routing protocols, including the novel application of green TE using SRv6, within the Lightspeed constellation. Our work introduces green TE and demonstrates its advantages over traditional protocols through comprehensive simulations and analysis.


\section{System Models and Methodology}
In this section, we present the system models utilized in this work, which include the simulation of Lightspeed constellation and the integration of SDN with LEO satellite networks using four routing protocols: SRv6, MPLS, IPv6, and IPv4, all based on OSPF, and the proposed green TE using SRv6. 

\subsection{System models}

\begin{figure}
    \centering
    \includegraphics[width=0.75\linewidth]{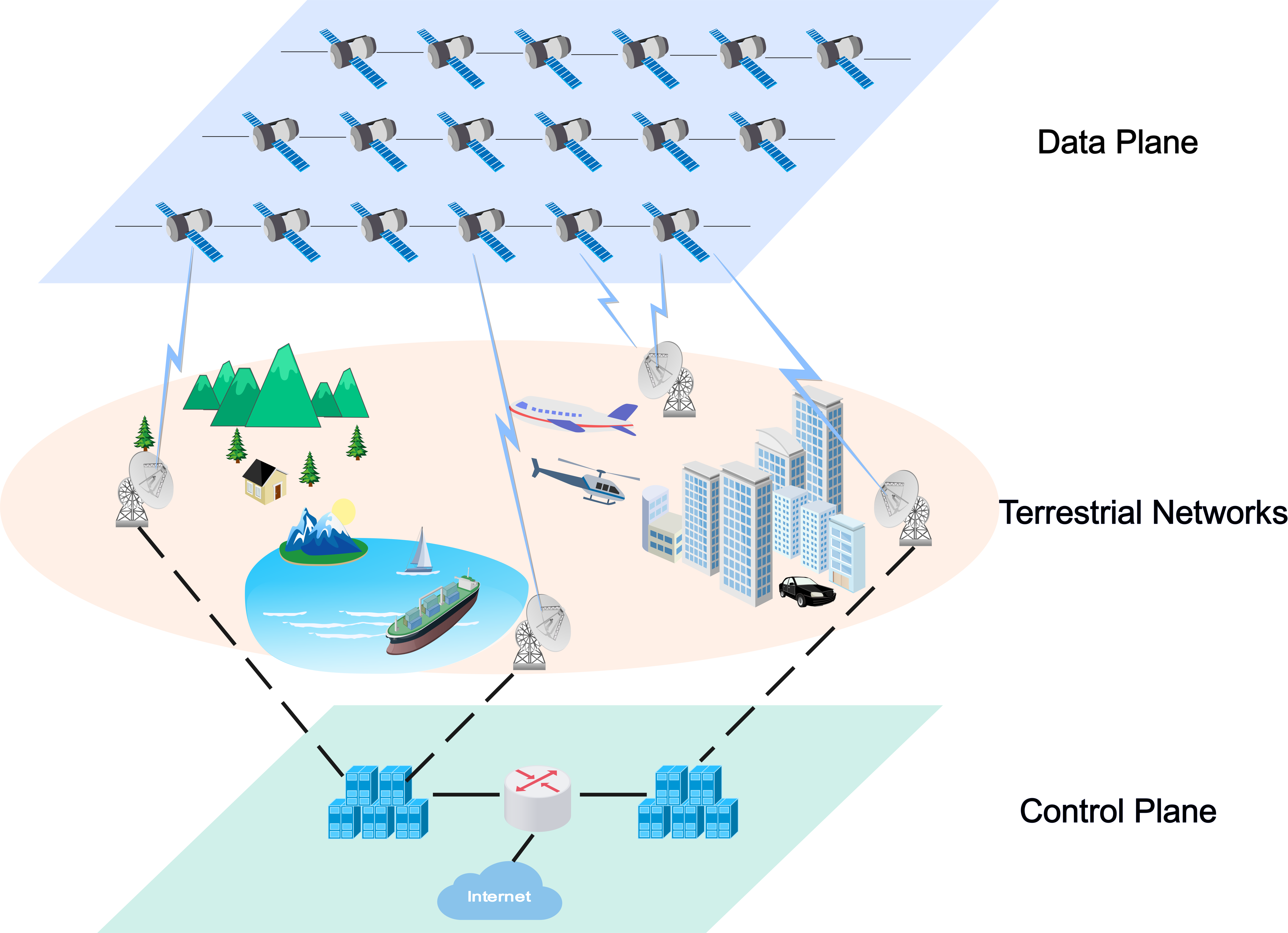}
    \caption{System architecture of SDN-based network.}
    \label{green}
     \vspace{-0.7cm}
\end{figure}

As shown in Fig. 1, the proposed system architecture is structured across three components: the data plane consists of LEO satellites and inter-satellite links (ISLs); the terrestrial network includes various users and ground stations; and the control plane uses the SDN controller. The satellites handle the forwarding of data packets between ground stations according to the SDN controller, which can dynamically adjust routes based on real-time network conditions and energy efficiency. The ground stations and user devices are scattered across diverse environments and also serve as intermediaries, relaying control data between the satellites and the controller.

We introduce a well-known and ongoing LEO satellite constellation, Lightspeed, which is comprised of 78 polar satellites orbiting at an altitude of 1,015 kilometer, and 120 inclined satellites at 1,325 kilometer, respectively. The constellation is engineered to deliver comprehensive global broadband internet services, including coverage of polar regions. 

In this work, satellites are assigned IPv6 addresses and configured to support SRv6. The flow tables on each satellite are programmed via OpenFlow 1.3 to match the active segment identifiers (SIDs) and perform necessary forwarding actions. Upon receiving a packet, a satellite processes the SR header by decrementing the segment field and updating the destination address to the next SID before forwarding. For traffic generation, we used iperf3 to craft and send packets with SR headers, allowing us to simulate and evaluate SRv6 routing performance under different traffic loads.

\begin{figure}
    \centering
    \includegraphics[width=0.8\linewidth]{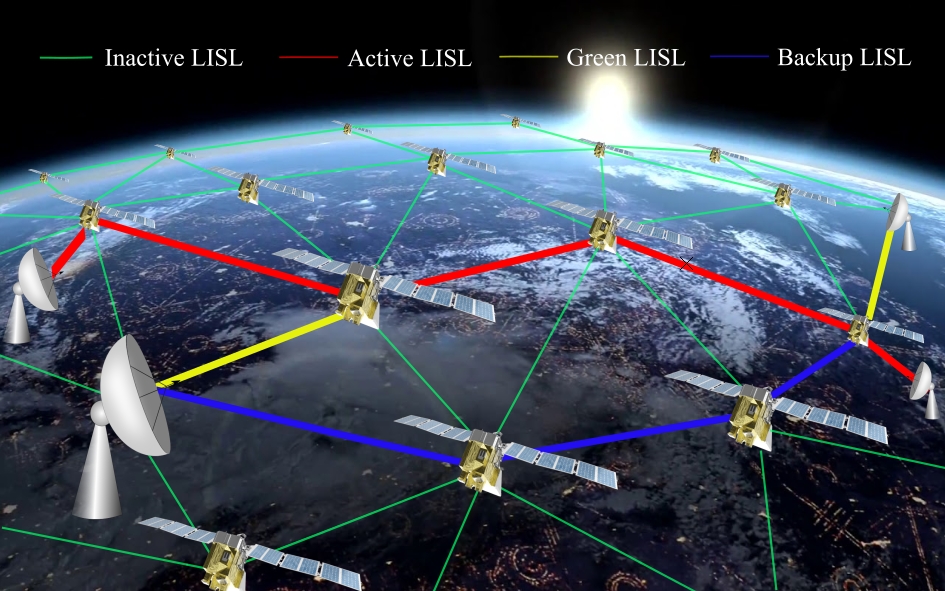}
    \caption{Green TE and backup rerouting path demonstration.}
    \label{green}
     \vspace{-0.7cm}
\end{figure}

Fig. 2 illustrates the green TE rerouting approach designed to minimize power consumption in the LEO satellite network. In this combined configuration, inactive ISLs are depicted in green, the red lines highlight the current active data path used for transmission, while the yellow lines indicate the green TE ISLs that operate in an energy-efficient manner by utilizing already active links and avoiding inactive satellites. Additionally, blue lines represent pre-configured backup paths that are not currently active but are ready to take over in case of disruptions or failures along the active path. This integrated approach ensures both energy efficiency and network reliability by balancing active resource utilization with redundancy for rerouting when necessary.

For MPLS, we extend the Ryu controller with MPLS functionalities. The controller computes label switched paths (LSPs) based on OSPF. MPLS label routing and stacking are handled within the controller. By matching on MPLS labels and configuring actions that emulate MPLS functionality, we simulate MPLS-like behavior using the Ryu controller. When a packet enters the network, the controller encapsulates it with an MPLS label stack for the computed LSP. As the packet traverses the network, each satellite swaps labels according to its forwarding information. 

In IPv4 and IPv6 implementations, we employ OSPF as the routing protocols within the SDN. The SDN controller configures each satellite to participate in OSPF routing, enabling them to exchange topology and link information dynamically. Each satellite computes the shortest paths to all other nodes using Dijkstra's algorithm and populates its routing table accordingly. Packet forwarding decisions are made based on the destination IP address present in the packet headers. However, since the four previously mentioned protocols are based on OSPF, the computed routes are expected to be identical.

\subsection{Proposed Green TE using SRv6}
This subsection introduces the proposed green TE using SRv6 to enhance energy efficiency in LEO satellite networks. The approach dynamically reroutes traffic from underutilized links to more active ones, allowing certain links to enter energy-saving mode and reducing overall energy consumption. By balancing traffic loads, it optimizes energy use while preserving network performance.

\begin{figure}
    \centering
    \includegraphics[width=1\linewidth]{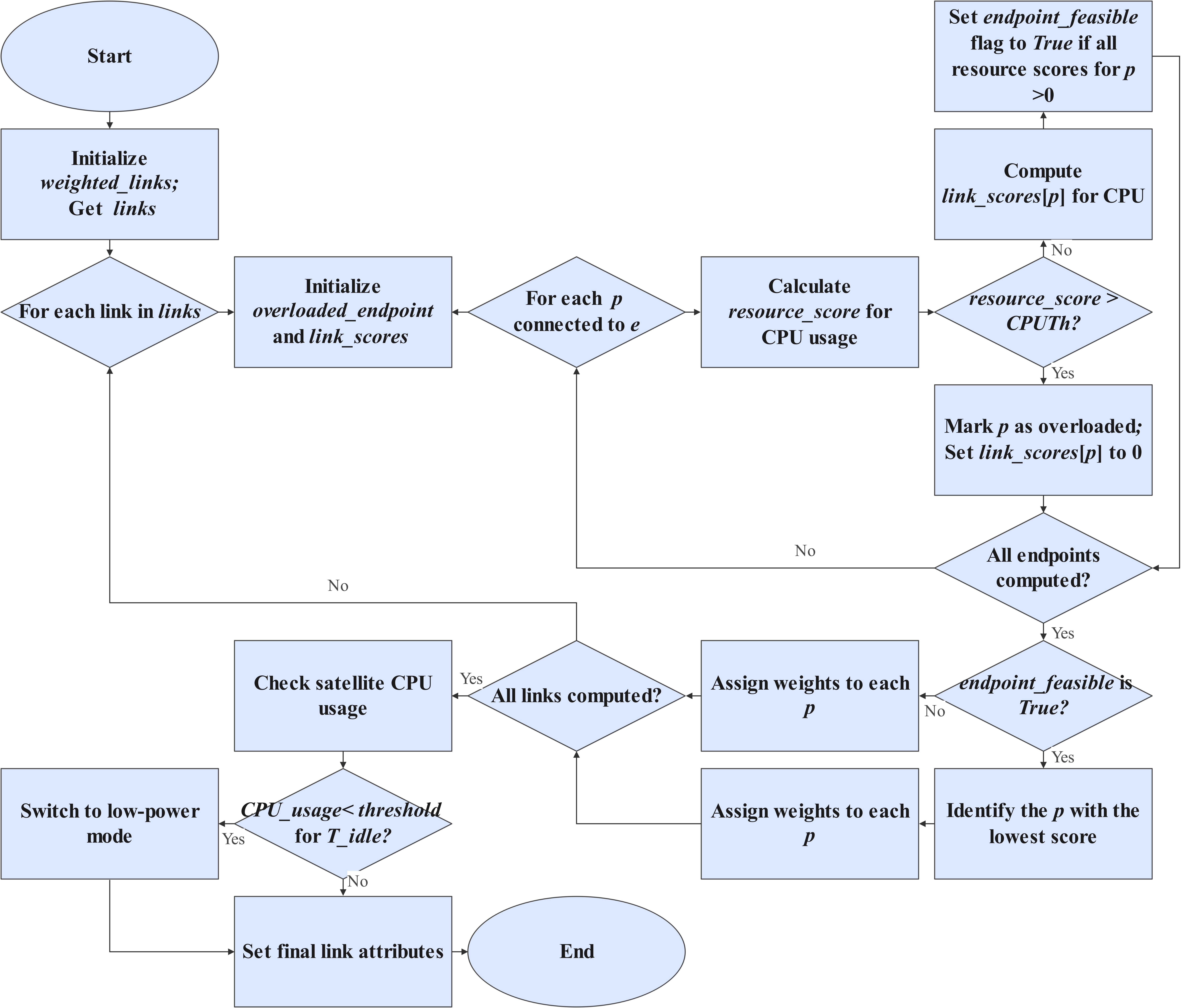}
    \caption{Flowchart demonstration of green TE using SRv6.}
    \label{green}
    \vspace{-0.7cm}
\end{figure}

To implement green TE, Algorithm 1, depicted in Fig. 3, continuously monitors CPU and memory usage on both ends of each link using \textit{psutil}. Based on these real-time CPU usage metrics, link weights are dynamically calculated using a modified weighted Dijkstra’s shortest path first (SPF) algorithm. The weight calculation involves subtracting the measured CPU from a baseline value to prioritize links with higher utilization. A Maximum Utilization Threshold \textit{CPUTh} of 80\% is imposed to prevent link overloads and reduce the likelihood of traffic loss, ensuring that a 20\% buffer remains available to handle potential traffic bursts \cite{jacob sr}. If the utilization of either endpoint of a link exceeds \textit{CPUTh}, the link is assigned a weight of zero, effectively removing it from the pool of viable routes for traffic forwarding. The updated link weights are then used in SRv6-based path computations, allowing the SDN controller to select the most energy-efficient routing paths. 

The algorithm optimizes resource usage and minimizes energy consumption by assigning links with higher utilization.  It emphasizes continuous resource monitoring and adaptive weight adjustments to dynamically respond to changing network conditions. Our main objective is to evaluate the energy usage and PDR of traditional routing protocols against the proposed green TE SRv6 rerouting approach. This comparison aims to demonstrate the effectiveness of the green TE in enhancing energy efficiency while maintaining robust and reliable network performance in LEO satellite constellations. No additional delay is considered in this work when implementing the green TE algorithm, as the controller proactively precomputes and frequently updates all routes. This ensures that no extra computation time is introduced on the satellites, maintaining real-time responsiveness and efficiency.

Meanwhile, the computational complexity of the proposed green TE algorithm based on OSPF is roughly \textit{O((M+N)logN)}, where \textit{M} is the number of links and \textit{N} is the number of endpoints on each link.  Each iteration computes resource scores, updates weights, and finds overloaded and idle satellites. Despite large propagation delays from ground controller to satellites, timely decisions are maintained since we keep updating the topology every 10 seconds and pre-compute backup and green routes. These localized updates reduce control overhead and allow incremental adjustments to ensure quick response. Moreover, the algorithm’s lightweight feature enables periodic updates without saturating long-delay links.

\vspace{-0.1cm}

\begin{algorithm}
\small  
\caption{Energy-efficient weight calculation algorithm for green SDN-based LEO satellite networks using SRv6}

\textbf{Inputs}:
Network object \textit{self} with links and resource metrics, CPU usage threshold \textit{CPUTh} for resource usage detection, Idle CPU threshold $\textit{CPU}_{\text{threshold}} = 10\%$, Idle time threshold $\textit{T}_{\text{idle}} = 10$ minutes.

\textbf{Output}:
Updated \textit{weighted\_links} with calculated link weights.

\textbf{BEGIN}

\begin{algorithmic}[1]
\State Initialize \textit{weighted\_links} dictionary for all links.
\State Get all links using \textit{get\_links()} and store in \textit{links}.

\For{each link $e$ in \textit{links}}
    \State Initialize \textit{overloaded\_endpoint} and \textit{link\_scores} for each endpoint in link $e$.
    
    \For{each endpoint $p$ connected to link $e$}
        \State Calculate \textit{resource\_score} for CPU usage.
        \If{\textit{resource\_score} exceeds \textit{CPUTh}}
            \State Add excess usage to \textit{overloaded\_endpoint\_scores[p]} and set \textit{link\_scores[p]} to $0$.
        \Else
            \State Compute \textit{link\_scores[p]} for CPU.
            \State Set \textit{endpoint\_feasible} flag to \textit{True} if all resource scores for $p$ are positive.
        \EndIf
    \EndFor
    
    \If{\textit{endpoint\_feasible} is \textit{True}}
        \For{each endpoint $p$ connected to link $e$}
            \State Set temporary weight for $p$ and update.
        \EndFor
    \Else
        \State Identify the endpoint with the lowest score.
        \For{each endpoint $p$ connected to link $e$}
            \State Set temporary weight for all endpoints on $e$.
        \EndFor
        \State Update \textit{weighted\_links} with actual score.
    \EndIf
\EndFor

\For{each satellite $s$ in \textit{self.satellites}}
    \State Monitor CPU usage of $s$: \textit{CPU\_usage(s)}.
    \If{\textit{CPU\_usage(s)} $<$ $\textit{CPU}_{\text{threshold}}$ for $\textit{T}_{\text{idle}}$}
        \State Switch satellite $s$ to low-power mode.
    \EndIf
\EndFor

\State Use \textit{nx.set\_link\_attributes} to set calculated link weights.
\end{algorithmic}

\textbf{END}
\end{algorithm}  

\vspace{-0.5cm}

\section{Numerical Results}
In this work, we simulate the Lightspeed LEO satellite constellation to investigate network performance across various routing algorithms under different traffic loads. Our analysis includes evaluating the peak and average CPU usage, average memory usage, and packet delivery rate of the satellites. Additionally, we examine packet overhead associated with each routing protocol.

\subsection{Simulation Setup}

\begin{figure}
    \centering
    \includegraphics[width=1\linewidth]{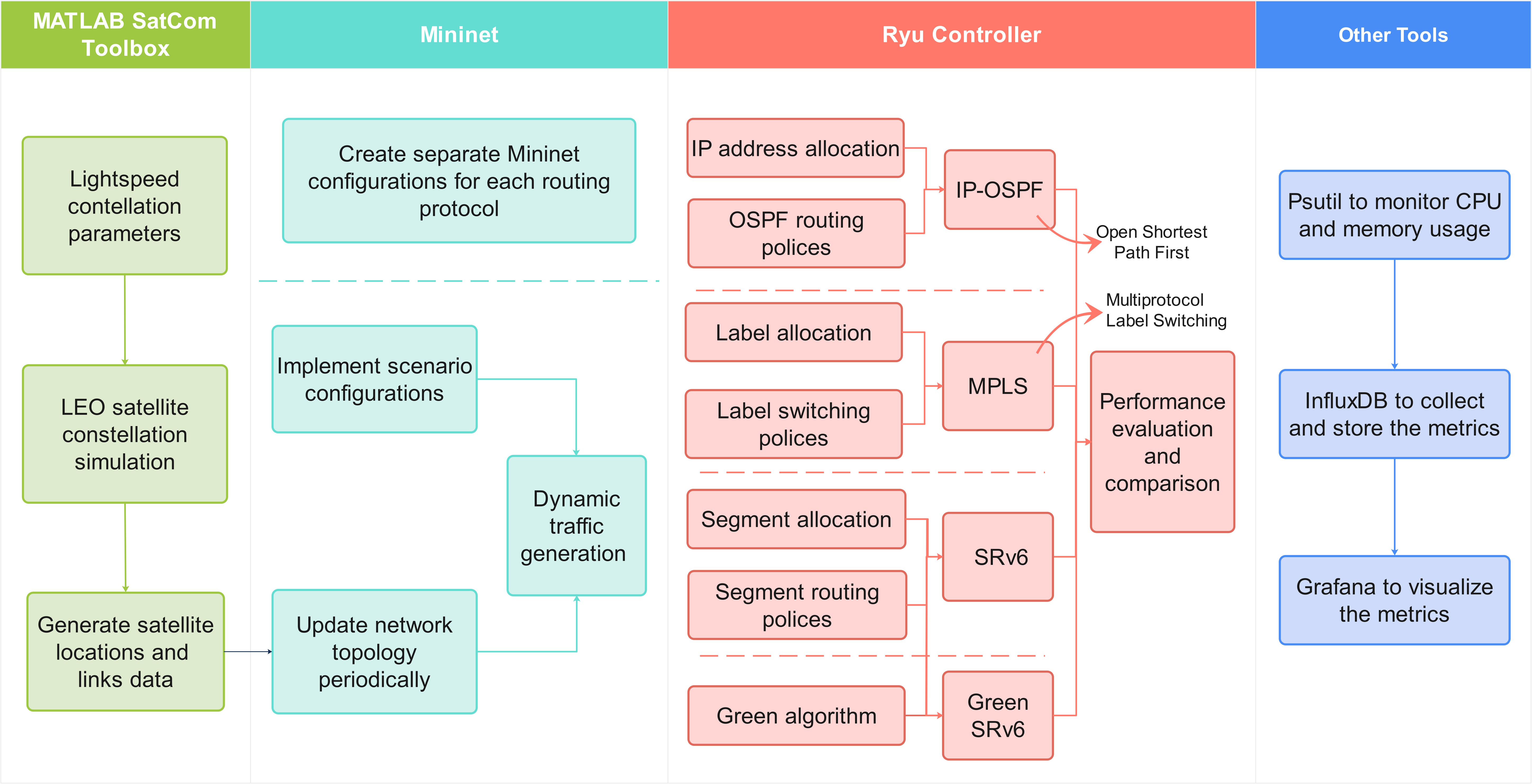}
    \caption{Simulation framework.}
    \label{tools}
     \vspace{-0.5cm}
\end{figure}


Fig. 4 illustrates the tools used to simulate the Telesat Lightspeed satellite constellation. Leveraging the well-known MATLAB Satellite Communication Toolbox, we propagate the satellite constellation and compute access intervals for both ISLs and satellite-to-ground station links. Additionally, we extract the time-evolving positions of satellites and ground stations and import this data into Mininet. To construct and update the network topology, we use NetworkX in Python, ensuring the topology is refreshed every 10 seconds in Mininet.

By integrating these tools and Python libraries, this work establishes a comprehensive simulation environment to investigate the performance of the green TE and other routing protocols. SDN provides the foundational architecture for dynamic network management, while Mininet and the Ryu controller facilitate the creation and control of virtual network topologies. InfluxDB and Grafana complement this setup by enabling resource monitoring and data visualization, respectively. The simulations and computations conducted for this paper are based on the Ubuntu virtual machine (VM) built upon the Linux open source system.

To study the performance of green TE using SRv6 and other routing protocols within the Lightspeed constellation, the period of the simulation is taken as 60 minutes. Table I shows the simulation parameters used in this work. We use CPU usage as a proxy for satellite power consumption because the scenario is simulated on a PC rather than actual routers. The results are inherently machine-dependent, influenced by hardware specifications and efficiency. A more accurate assessment would require energy profiling tools or power models that account for actual satellite hardware characteristics.

\begin{table}
\centering
\caption{Simulation Parameters.}
\label{table1}
\setlength{\tabcolsep}{12pt}
\arrayrulecolor{black}
\begin{tabular}{ll} 
\hline
\textbf{Parameter} & \textbf{Value} \\ 
\hline
Number of satellites & 198 \\
Number of ground stations & 10 \\
Number of SDN controllers & 2 \\
Max traffic capacity & 20 Mbps \\
Dynamic topology update frequency & 10 sec \\
Packet payload size & 512 Bytes \\
CPU maximum usage threshold & 80\% \\
Traffic packet timeout & 1 sec \\
\hline
\end{tabular}
\arrayrulecolor{black}
\vspace{-0.5cm}
\end{table}

\subsection{Network Performance Results}
The numerical results for network performance metrics, including peak and average CPU usage, average memory usage, and average PDR, are presented in Fig. 5–Fig. 8.

\begin{figure}[!b]
\centering
\vspace{-0.8 cm}
\includegraphics[width=0.9\linewidth]{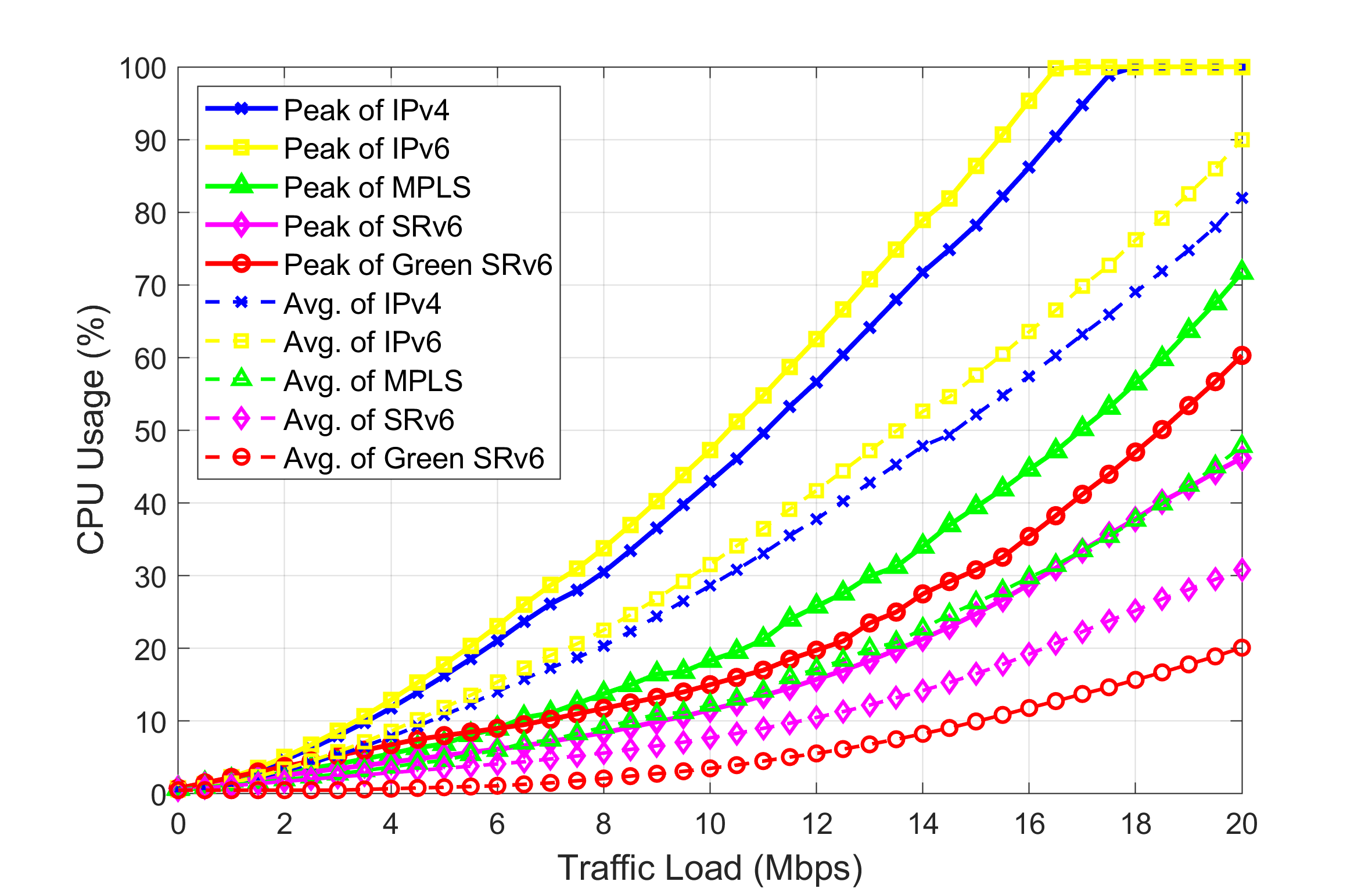}
\caption{Peak and average CPU usage vs. traffic load for different routing protocols.}
\label{fig:cpu_usage}
\vspace{-0.5 cm}
\end{figure}

\begin{figure}[!b]
\centering
\includegraphics[width=0.9\linewidth]{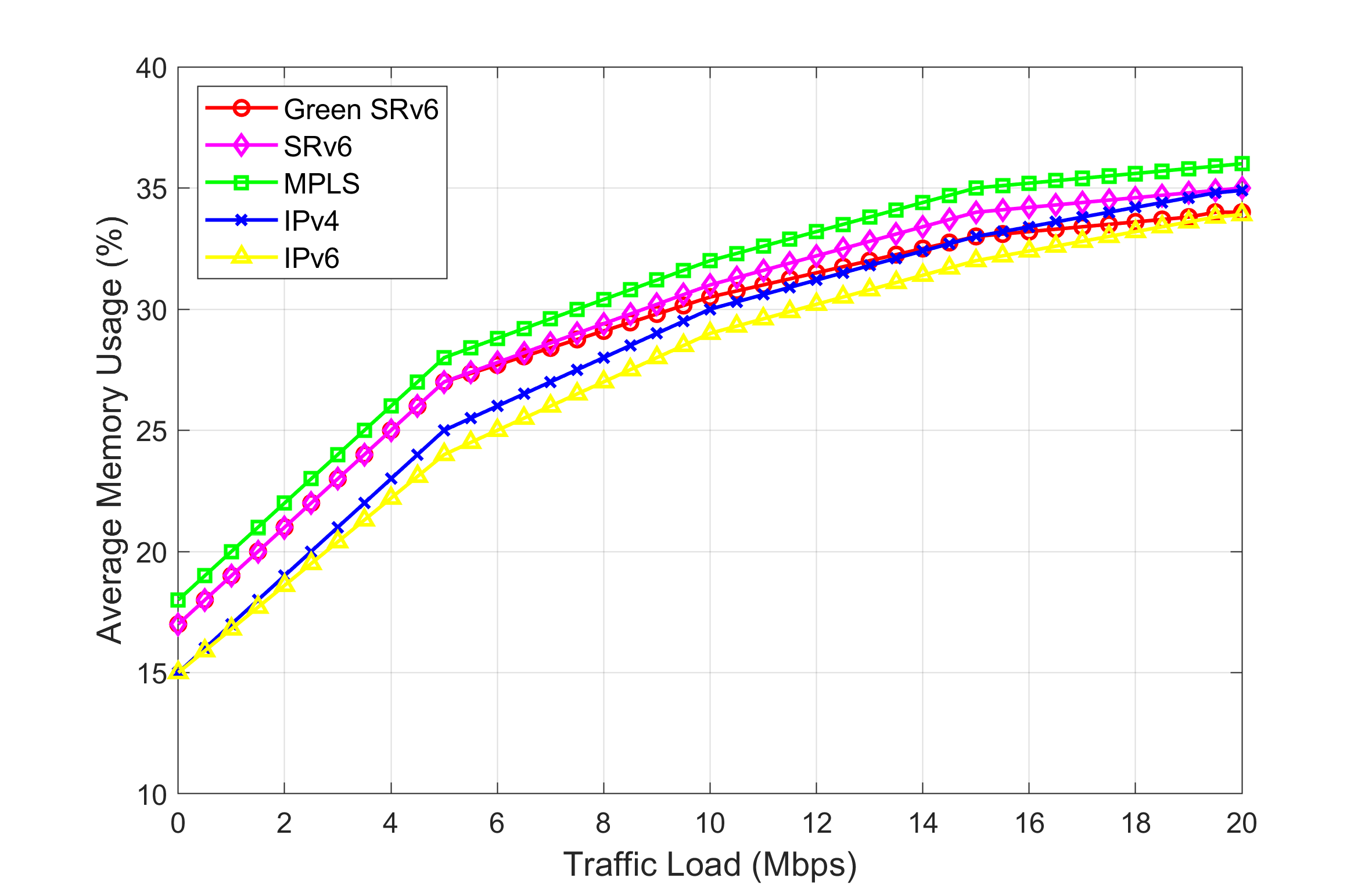}
\caption{Average memory usage vs. traffic load for different routing protocols.}
\label{fig:memory_usage}
\end{figure}

\begin{figure}[!b]
\centering
\vspace{-0.5 cm}
\includegraphics[width=0.9\linewidth]{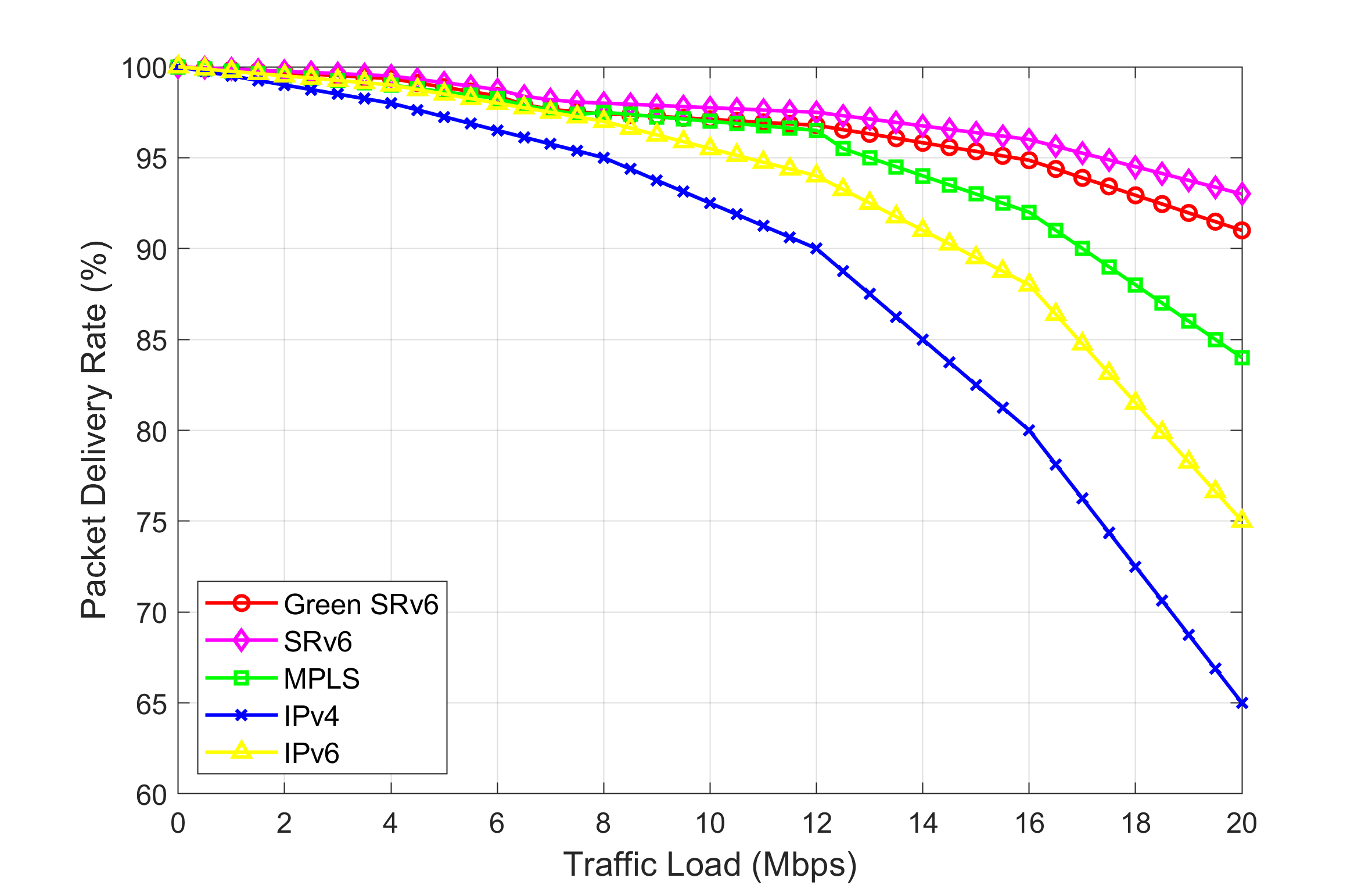}
\caption{Average packet delivery rate vs. traffic load for different routing protocols.}
\label{fig:pdr}
\end{figure}

\begin{figure}[!b]
\centering
\vspace{-0.5 cm}
\includegraphics[width=0.9\linewidth]{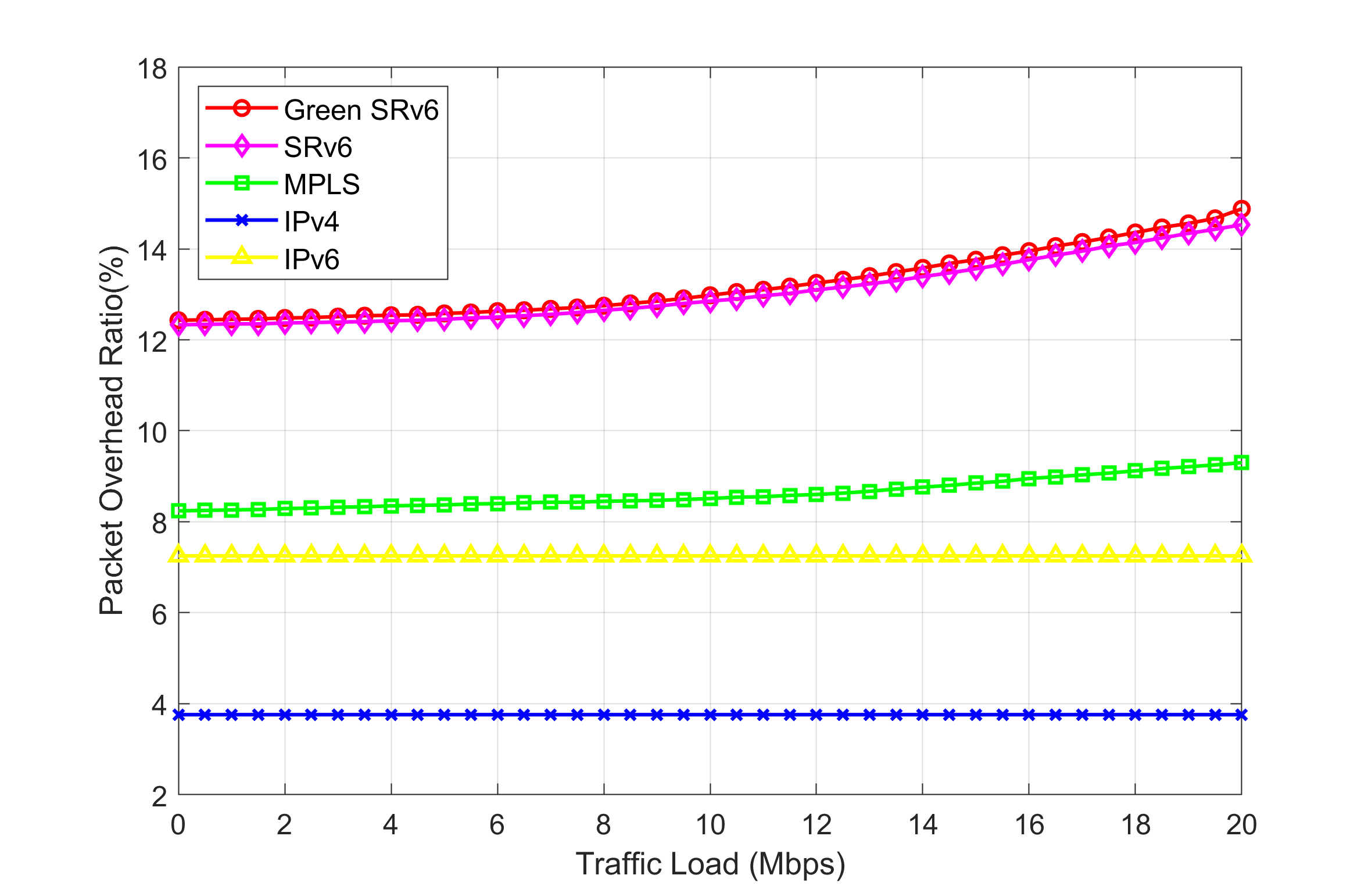}
\caption{Average packet overhead ratio vs. traffic load for different routing protocols.}
\label{fig:overhead}
\end{figure}

In Fig. 5, both peak (shown as dashed lines) and average (shown in solid lines) CPU usage increase as the traffic load rises for all routing protocols, reflecting the growing processing demands on satellites as traffic load increases. IPv4 (shown in blue lines) and IPv6 (shown in yellow lines) exhibit significantly higher CPU usage comparing to MPLS (shown in green lines) and SRv6. Both IPv4 and IPv6 can reach close to 100\% peak CPU usage under maximum traffic load. In contrast, SRv6, particularly the proposed Green SRv6 (shown in red lines), demonstrates much lower average CPU usage. This is largely because SRv6 minimizes the need for route recalculations when the topology changes, unlike traditional protocols, which need to wait for periodic updates to maintain routing and topology accuracy. The waiting time in traditional protocols correlates with higher CPU usage, likely because each update triggers a route recalculation. Regarding peak CPU usage, green SRv6 exhibits slightly higher values than the original SRv6 (shown in purple lines), as this increase results from the rerouting algorithm designed for energy efficiency.

In Fig 6, average memory usage also increases with the traffic load for all protocols but at a much slower rate compared to average CPU usage. The differences between protocols are less pronounced, but MPLS generally shows the highest memory usage, followed by green SRv6, the original SRv6 and IPv4, with IPv6 demonstrating the lowest usage. The gradual increase and leveling off at higher traffic loads indicate that memory usage is relatively stable, with limited additional overhead as traffic approaches full capacity. This stability reflects memory management efficiencies within each protocol, with green SRv6 showing slight drawbacks.

Fig. 7 shows a decrease in PDR as traffic load increases, with IPv4 experiencing the steepest drop-off, reaching below 70\% PDR at full capacity. The proposed green SRv6, original SRv6 maintain higher PDRs throughout the range even under high load. This suggests that SRv6 and MPLS handle packet forwarding more effectively under congestion, likely due to their efficient path management capabilities, whereas IPv4 and IPv6 struggle to maintain delivery rates at higher loads and can drop packets while waiting for topology updates.

Fig. 8 shows the results for packet overhead. We see a consistent pattern across different routing algorithms as traffic load increases, with each protocol maintaining a nearly stable overhead percentage. The original SRv6 and the green SRv6 show the highest packet overhead at around 12\%, followed by MPLS at approximately 8\%, IPv6 around 7\%, and IPv4 at around 4\%. SRv6 and MPLS, which are designed with advanced routing features, inherently carry more overhead to support these functionalities, and slightly increase with high traffic load due to the backup path, while IPv4, being the simplest protocol, has the least packet overhead. 

These three figures collectively highlight the trade-offs between different routing protocols in terms of peak and average CPU usage, average memory usage, as well as PDR under varying traffic loads. The proposed green TE using SRv6, appears to be the most balanced protocol, providing significantly lower average CPU usage while maintaining a relatively high PDR under heavy load conditions. Traditional TE like IPv4 and IPv6 based on OSPF, in contrast, exhibits higher peak and average CPU usage and a more significant drop in PDR, making it less suitable for high-demand scenarios. These findings suggest that the proposed green SRv6, with its optimized rerouting and energy-efficient algorithm, is a promising candidate for SDN-based LEO satellite networks, especially when energy efficiency and reliability are priorities.

\section{Conclusions}

In this paper, we proposed a green TE solution using SRv6 and investigated the performance of different routing protocols with varying traffic loads. Based on simulations, the integration of green TE with SRv6 in the Lightspeed constellation shows promising results. The results reveal that green SRv6 maintains lower average CPU usage than traditional protocols, achieving significant energy efficiency improvements. Additionally, green SRv6 sustains a high PDR under increasing traffic loads, supporting robust and reliable communications in LEO networks. Memory usage remains relatively stable across protocols, with SRv6 and MPLS consuming slightly more resources due to their advanced routing features.


Future work will focus on further optimizing green TE to enhance energy efficiency in LEO satellite constellations. A predictive, energy-aware green TE approach using machine learning for traffic forecasting is expected to enable proactive rerouting, improving efficiency. Additionally, a deeper comparative analysis of SRv6 against other emerging protocols is set to provide valuable insights into optimizing energy usage in next-generation LEO satellite networks, particularly as global connectivity demands continue to rise.

\section*{Acknowledgment}
This work has been supported by the National Research Council Canada
(NRC), MDA, Mitacs, and Defence R\&D Canada (DRDC), within the Optical
Satellite Communications Consortium (OSC) framework in response to the
High Throughput Secure Networks (HTSN) challenge program of the
Government of Canada.

\end{document}